%% file: main.tex
\pdfoutput=1

\documentclass[11pt]{article}

\usepackage{acl}
\usepackage{times}
\usepackage{latexsym}
\usepackage{listings}
\usepackage{xcolor}

\usepackage[T1]{fontenc}

\usepackage[utf8]{inputenc}

\usepackage{microtype}

\usepackage{booktabs}
\usepackage{graphicx}

\definecolor{codegreen}{rgb}{0,0.6,0}
\definecolor{codegray}{rgb}{0.5,0.5,0.5}
\definecolor{codepurple}{rgb}{0.58,0,0.82}
\definecolor{backcolour}{rgb}{1,1,1}
\lstdefinestyle{mystyle}{
  backgroundcolor=\color{backcolour}, commentstyle=\color{codegreen},
  keywordstyle=\color{magenta},
  numberstyle=\tiny\color{codegray},
  stringstyle=\color{codepurple},
  basicstyle=\ttfamily\footnotesize,
  breakatwhitespace=false,         
  breaklines=true,                 
  captionpos=b,                    
  keepspaces=true,                 
  numbers=none,                    
  numbersep=5pt,                  
  showspaces=false,                
  showstringspaces=false,
  showtabs=false,                  
  tabsize=2
}

\title{CyNER: A Python Library for Cybersecurity Named Entity Recognition
}

\author{Md Tanvirul Alam$^1$, Dipkamal Bhusal$^1$, Youngja Park$^2$, Nidhi Rastogi$^1$\\
  $^1$Rochester Institute of Technology, USA \\
  \texttt{\{ma8235,db1702,nxrvse\}@rit.edu}\\
    $^2$IBM T. J. Watson Research Center, USA \\
  \texttt{young\_park@us.ibm.com}\\
 }

\begin{document}
\maketitle
\begin{abstract}
Open Cyber threat intelligence (OpenCTI) information is available in an unstructured format from heterogeneous sources on the Internet. We present CyNER, an open-source python library for cybersecurity named entity recognition (NER). CyNER combines transformer-based models for extracting cybersecurity-related entities, heuristics for extracting different indicators of compromise, and publicly available NER models for generic entity types. We provide models trained on a diverse corpus that users can readily use. Events are described as classes in previous research - MALOnt2.0~\cite{christian2021ontology} and MALOnt~\cite{rastogi2020malont} and together extract a wide range of malware attack details from a threat intelligence corpus. The user can combine predictions from multiple different approaches to suit their needs. The library is available in Github \footnote{https://github.com/aiforsec/CyNER} and a video demonstrating the use case is available in YouTube.\footnote{https://youtu.be/E4uKCrKKaP8}
\end{abstract}

\section{Introduction}
A large amount of open cyber threat and attack information, also called cyber threat intelligence (openCTI), is available on the Internet on platforms such as security blogs, the dark web, software vendors bulletin boards, official news and social networks~\cite{yi2020cybersecurity}. Both structured and unstructured forms of openCTI can benefit security operations center (SOC) analysts and researchers in their increasingly complex jobs of triaging false alerts, finding zero-day attacks, protecting intellectual property, and preventing intrusions from adversaries. However, access to this information from openCTI has remained relatively untapped due to limited research enabling aggregation and structuring of threat intelligence incidents. Furthermore, there is a lack of a gold standard dataset in cybersecurity, which is the first prerequisite for the basic task of information extraction or train any modern natural language processing (NLP) method.
\par
Current methods rely on structured knowledge bases such as the cybersecurity vulnerability enumeration (CVE)\footnote{https://cve.org} database, mitre att\&ck framework(att\&ck)\footnote{https://attack.mitre.org/} to extract named entities which result in high accuracy but produce poor performance when similar entities are extracted from unstructured, complex threat reports, which is more often the case. Conditional random fields (CRF)~\cite{joshi2013extracting}, support vector machine (SVM)~\cite{deliu2017extracting},  and many learning models have been unsuccessful in yielding satisfactory results in extracting cybersecurity events from openCTI. While transformer-based methods, such as BERT~\cite{devlin2018bert}, have been explored, the cybersecurity concepts extracted provided a limited advantage over information extracted from structured sources such as CVE ~\cite{xie2021named}.

\par
To address the challenges above,  we propose \textsf{CyNER}, a generic python library for extracting cybersecurity threat intelligence incidents. Specifically, we make the following contributions:
\begin{enumerate}
    \item We provide a manually-labeled, benchmark dataset annotated on a wide range of cybersecurity incidents. While the annotated corpus relates to Android malware threat analysis, the concepts are generic enough to apply to other security threats. We also provide a corpus of noise-free, high-quality openCTI reports written on android malware analysis.
    \item We provide \textsf{CyNER}, an easy to use python library that allows access to transformer-based NER models pre-trained on high-fidelity security events (e.g., malware, vulnerability) and a heuristic model for threat indicators (e.g., malware hashes, compromised IP addresses). In addition, we include support for extracting generic entity types with popular NLP libraries.
    \item We provide a flexible and modular implementation of CyNER and user documentation. The library provides the option to combine prediction from multiple models with configurable priority settings.
\end{enumerate}
Our proposed framework, \textsf{CyNER} allows cybersecurity information extraction using NER models that deliver high accuracy. The malware-related threat intelligence discovery combines malware data with general world concepts, integrates "knowledge," correlation, and concept grouping, and therefore, generates inference and insights for SOC analysts.
\par
In this work, we instantiate existing malware ontologies~\cite{rastogi2020malont, christian2021ontology} using a corpus comprising over 60 high quality and large threat intelligence reports written between {2018-2021} and downloaded from websites belonging to security and technology organizations~(Kaspersky, Symantec, McAfee). The annotated text contains entities manually generated by graduate students trained by their advisor experienced in cybersecurity. We build on the existing strengths of the security, semantic web, and natural language domains. Security ontologies \cite{rastogi2020malont} \cite{swimmer} are extended to extract cyber security threat intelligence from online sources. Other benefits include the interoperability of information across applications and creating labeled features for machine learning models. Using ontologies when labeled data allows queries over large datasets to identify trends and predict future occurrences.
\section{Related Work}
Security ontologies (like malware) have attempted to provide a structure to security concepts that capture event analysis~\cite{swimmer} primarily to collate data and distribute it across organizations and institutions~\cite{stix,connolly2014trusted}. Other models for extracting and structuring threat intelligence focus on one of the following - identifying attack patterns, software vulnerabilities, threats, or information dissemination.
\par
CVE\footnote{https://cve.mitre.org/}, NVD\footnote{https://nvd.nist.gov/} are vulnerability tracking programs where the security community feeds information. For sharing threat intelligence, there are industry standards like Structured Threat Information eXpression (STIX)\cite{stix}, which provides a language-agnostic framework to capture threat intelligence into a shareable package. In contrast, a trusted automated exchange of indicator information (TAXII) \cite{connolly2014trusted} is a platform that can send and receive the STIX package.

\input{sections/system}

\input{sections/datasets}
\input{sections/evaluation}
\input{sections/demo}

\input{sections/conclusion}


\bibliography{anthology}

\end{document}

%% file: sections/system.tex
\section{System Overview}

\begin{figure*}[t]
\centering
\includegraphics[width=1\textwidth]{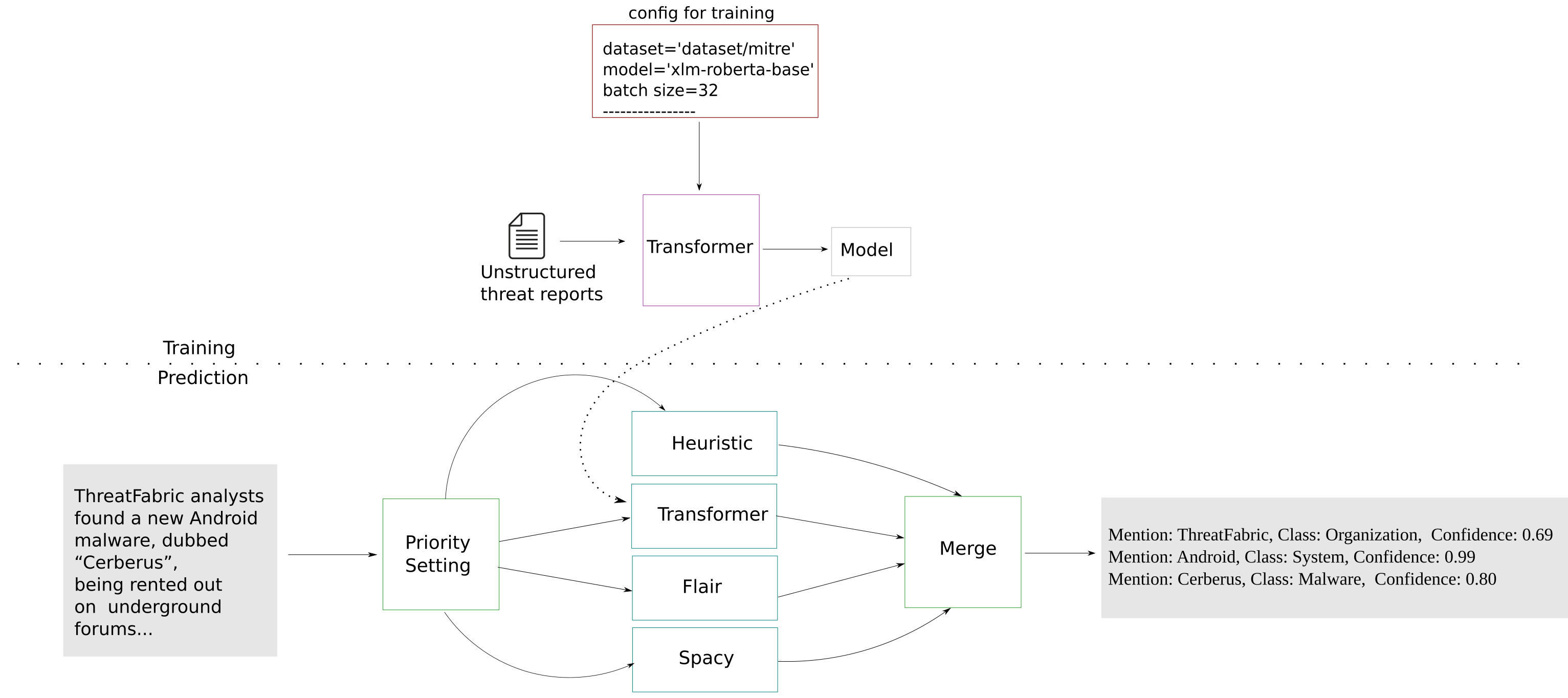}
\caption{System diagram of the CyNER library}
\label{fig:system}
\end{figure*}

The primary goal of our library is to provide cybersecurity researchers with models for extracting cybersecurity entities and provide the framework to train new models on annotated datasets. We show a detailed system diagram of our library in Figure \ref{fig:system}. In this section, we discuss the three key components of our framework - using pretrained models for prediction, training models on new datasets, and augmenting prediction using a combination of different models.

\subsection{Pretrained Model for Prediction}
We provide three different types of models that users can readily use to extract cybersecurity entities-
\subsubsection{Transformer models pretrained on cybersecurity corpus} We train transformer models on an annotated cybersecurity corpus and add the pretrained models in our library. Our dataset consists of five classes-malware, indicator, system, organization, and vulnerability. These models allow users to extract named entities related to the cybersecurity domain. We use the \textit{tner} library \cite{ushio2021t} to train the models. 

\subsubsection{Heuristic for indicators of compromise} We provide a heuristic-based approach for categorizing different indicators of compromise, e.g., URL, IP address, hash, etc. These entities do not require an understanding of the context they appear in and can be usually extracted with regular expression matching. We use regular expression templates proposed in earlier work on cybersecurity NER \cite{yi2020cybersecurity,piplai2020creating}. We show some example regex patterns in Table \ref{table:regex}.

\subsubsection{Generic NER models}
Sometimes entities that do not fall under cybersecurity concepts may be of interest when analyzing threat reports. For example, extracting targeted countries of a malware attack from threat reports can give important insight regarding its behavior. Off-the-shelf NER models provided in popular NLP libraries, which are trained on generic corpus like Ontonotes \cite{hovy2006ontonotes} can achieve good accuracy on such classes of entities \cite{kim2020automatic}. We integrate Flair \cite{akbik2019flair} and spaCy\footnote{https://spacy.io/} with our library, and users can readily use the pretrained NER models available in those libraries.

\subsection{Model Training} 
We provide modules for finetuning transformer language models on user-provided datasets. We use Huggingface's transformer library \cite{wolf2019huggingface} for finetuning. The user can use pretrained language models available in the transformer library and set different hyperparameters during model finetuning. 

\subsection{Combining Prediction from Multiple Approaches} We provide a simple yet intuitive mechanism for combining predictions from different methods. The user can use any combination of the available models (transformers, heuristics, flair, spacy) for making a prediction. The user can also define priority when merging outputs from multiple models using a keyword argument \textit{priority}. The default value of the \textit{priority} is  'HTFS,' which indicates that the output from the heuristic approach will be given the highest priority, followed by the prediction from the given transformer model, followed by the Flair and Spacy models (if provided). If there is an overlap between predicted entity positions from multiple approaches, we use the prediction from the highest priority one.

\begin{table}[]
\centering
\resizebox{\linewidth}{!}{%
\begin{tabular}{@{}ll@{}}
\toprule
Entity Types & Regular Expression                                                                                                                                           \\ \midrule
FilePath     & {}r'{[}a-zA-Z{]}:\textbackslash{}\textbackslash{}({[}0-9a-zA-Z{]}+)', r'(\textbackslash{}/{[}\textasciicircum{}\textbackslash{}s\textbackslash{}n{]}+)+'{} \\
Email        & {}r'{[}a-z{]}{[}\_a-z0-9-.{]}+@{[}a-z0-9-{]}+{[}a-z{]}+'{}                                                                                                                                                                      \\
SHA256       & {}r'{[}a-f0-9{]}\{64\}|{[}A-F0-9{]}\{64\}'{}                                                                               \\
SHA1         & {}r'{[}a-f0-9{]}\{40\}|{[}A-F0-9{]}\{40\}'{}                                                                                                                                         \\
CVE          & {}r'CVE—{[}0-9{]}\{4\}—{[}0-9{]}\{4,6\}'{}                                                                                                                 \\
IPv4         & {}r'\textasciicircum{}((25{[}0-5{]}|(2{[}0-4{]}|1\textbackslash{}d|{[}1-9{]}|)\textbackslash{}d)(\textbackslash{}.(?!$)|$))\{4\}\$'{}                \\\bottomrule     
\end{tabular}%
}
\caption{Example regular expressions used in heuristics based detection}
\label{table:regex}
\end{table}

%% file: sections/datasets.tex
\section{Datasets}

\begin{figure*}[h]
\centering
\includegraphics[width=1\textwidth]{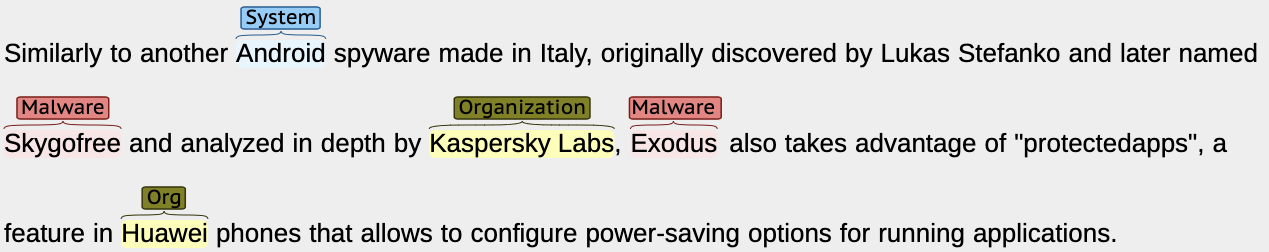}
\caption{Example annotation from BRAT}
\label{fig:brat}
\end{figure*}

We collected approximately 60 threat intelligence reports referenced in the MITRE att\&ck website under the software category.\footnote{https://attack.mitre.org/software/} Each threat report describes a unique malware. Since cleaning unstructured CTI reports can be very challenging \cite{kim2020automatic}, we manually extract the clean text from each report. We use the BRAT annotation tool \cite{stenetorp2012brat} for annotating the corpus. Example annotation is shown in Figure ~\ref{fig:brat}.


Our dataset consists of five classes of high importance for analyzing threat reports - malware, indicators of compromise, system, organization, and vulnerability. The malware class encapsulates viruses, trojans, ransomware, etc. Operating system (e.g., Android, Windows), software (e.g., Adobe flash player, Skype), and hardware. Indicators of compromise comprise domain name, URL, IP address, filename, Hash, email, port number, etc. Vulnerability includes both CVE ID (e.g., CVE-2012-2825) and mention of exploits (e.g., master key vulnerability). It is worth noting that other types of entities like location and person are also important when analyzing threat reports. However, these and other generic entities can be extracted well using different off-the-shelf models \cite{kim2020automatic}. As a result, we added the support to extract such entities using other NLP libraries and did not include them in our annotation.

We split the corpus into train, dev, and test split containing 40, 10, and 10 documents, respectively. Statistics of different types of entities in each split are displayed in Table \ref{tab:dataset}. We convert the annotation in BIO format for training, B-prefix indicating the beginning of a tag and the I-prefix indicating the inside of that tag. An outside (O) tag is used to indicate a token that does not belong to any of the five categories. We have 106991 total tokens in the complete corpus, with 4530 tagged entities.

\begin{table}[t]
\centering
\resizebox{\linewidth}{!}{%
\begin{tabular}{@{}lccccc@{}}
\toprule
\textbf{Split} & \multicolumn{1}{l}{\textbf{Malware}} & \multicolumn{1}{l}{\textbf{Indicator}} & \multicolumn{1}{l}{\textbf{System}} & \multicolumn{1}{l}{\textbf{Organization}} & \multicolumn{1}{l}{\textbf{Vulnerability}} \\ \midrule
Train & 703 & 1021 & 837 & 284 & 48 \\
Dev & 254 & 208 & 182 & 92 & 9 \\
Test & 242 & 261 & 248 & 131 & 10 \\ \bottomrule
\end{tabular}%
}
\caption{Dataset statistics showing the number of instances of different classes.}
\label{tab:dataset}
\end{table}

%% file: sections/evaluation.tex
\section{Evaluation}
\subsection{Experimental Settings}
We finetune transformer models available in Huggingface's transformers library \cite{wolf2019huggingface} for sequence tagging. We train both base and large variants of BERT \cite{devlin2018bert}, RoBERTa \cite{liu2019roberta} and XLM-RoBERTa \cite{conneau2019unsupervised} models. We add a linear layer to the hidden layer representation obtained from the transformer model for the token classification task.
We train each model for 20 epochs with a sequence length of 128. We use an initial learning rate of 1e-5 for the base models and 5e-6 for the large models. We use 32 samples for each mini-batch. All models are trained using AdamW optimizer \cite{loshchilov2017decoupled} on a single Nvidia Tesla V100 GPU.

\subsection{Results}
We report the span micro-F1 score computed by the seqeval \footnote{https://pypi.org/project/seqeval/} library. We report results for the six different transformer models on Table \ref{tab:result}. XLM models outperform both BERT and RoBERTa models, despite being multilingual. XLM-RoBERTa-large model achieves the best average F1 score of 76.66\%. Class-specific precision, recall, and F1-score are displayed in Table \ref{tab:class-result}. The model performs relatively poorly on the Organization class compared to others.

\begin{table}[]
\centering
\resizebox{0.95\linewidth}{!}{%
\begin{tabular}{@{}lccc@{}}
\toprule
\textbf{Model} & \textbf{Precision} & \textbf{Recall} & \textbf{F1-score} \\ \midrule
BERT-base-uncased & 69.67 & 69.88 & 69.77 \\
BERT-large-uncased & 72.69 & 73.45 & 73.07 \\
RoBERTa-base & 37.22 & 42.50 & 39.69 \\
RoBERTa-large & 34.76 & 44.18 & 38.91 \\
XLM-RoBERTa-base & 74.57 & 77.23 & 75.88 \\
XLM-RoBERTa-large & \textbf{75.30} & \textbf{78.07} & \textbf{76.66} \\
\bottomrule
\end{tabular}%
}
\caption{Result on the test dataset for different Transformer models.}
\label{tab:result}
\end{table}

\begin{table}[]
\centering
\resizebox{0.9\linewidth}{!}{%
\begin{tabular}{@{}lccc@{}}
\toprule
\textbf{Class} & \textbf{Precision} & \textbf{Recall} & \textbf{F1-score} \\ \midrule
Malware & 79.82 & 75.11 & 77.39 \\
Indicator & 78.34 & 86.62 & 82.27 \\
System & 70.36 & 79.93 & 74.84 \\
Organization & 70.64 & 60.16 & 64.98 \\
Vulnerability & 100.0 & 80.0 & 88.89\\
\bottomrule
\end{tabular}%
}
\caption{Result for different classes in test set for XLM-RoBERTa-large model.}
\label{tab:class-result}
\end{table}

%% file: sections/demo.tex
\begin{lstlisting}[float=*,frame=single,label={lst:1}, language=Python, caption={Prediction with only pretrained transformer model (Mention: entity extracted from the text, Class: entity label, Start: starting character position in the given text, End: ending character position in the given text, Confidence: confidence/probability of detection)}]
import cyner

text='Proofpoint report mentions that the German-language messages were turned off once the UK messages were established, indicating a conscious effort to spread FluBot 446833e3f8b04d4c3c2d2288e456328266524e396adbfeba3769d00727481e80 in Android phones.'

model = cyner.CyNER('xlm-roberta-large', use_heuristic=False, flair_model=None)

entities = model.get_entities(text)
for e in entities:
    print(e)

#output
Mention: Proofpoint, Class: Organization, Start: 0, End: 10, Confidence: 0.82
Mention: FluBot, Class: Malware, Start: 156, End: 162, Confidence: 0.92
Mention: 446833e3f8b04d4c3c2d2288e456328266524e396adbfeba3769d00727481e80, Class: Indicator, Start: 163, End: 227, Confidence: 0.90
Mention: Android, Class: System, Start: 231, End: 238, Confidence: 1.00
\end{lstlisting}

\begin{lstlisting}[float=*,frame=single,label={lst:2},language=Python, caption=Prediction with a pretrained transformer model and heuristic for indicators]
model = cyner.CyNER(transformer_model='xlm-roberta-large', use_heuristic=True, flair_model=None, priority='HTFS')

entities = model.get_entities(text)
for e in entities:
    print(e)

#output
Mention: 446833e3f8b04d4c3c2d2288e456328266524e396adbfeba3769d00727481e80, Class: SHA256, Start: 163, End: 227, Confidence: 1.00
Mention: Proofpoint, Class: Organization, Start: 0, End: 10, Confidence: 0.82
Mention: FluBot, Class: Malware, Start: 156, End: 162, Confidence: 0.92
Mention: Android, Class: System, Start: 231, End: 238, Confidence: 1.00
\end{lstlisting}

\begin{lstlisting}[float=*,frame=single,label={lst:3},language=Python, caption={Prediction using pretrained transformer model, heuristic for indicators and Flair NER model}]
model = cyner.CyNER(transformer_model='xlm-roberta-large', use_heuristic=True, flair_model='ner', priority='HTFS')

entities = model.get_entities(text)
for e in entities:
    print(e)

#output
Mention: 446833e3f8b04d4c3c2d2288e456328266524e396adbfeba3769d00727481e80, Class: SHA256, Start: 163, End: 227, Confidence: 1.00
Mention: Proofpoint, Class: Organization, Start: 0, End: 10, Confidence: 0.82
Mention: FluBot, Class: Malware, Start: 156, End: 162, Confidence: 0.92
Mention: Android, Class: System, Start: 231, End: 238, Confidence: 1.00
Mention: German-language, Class: MISC, Start: 36, End: 51, Confidence: 1.00
Mention: UK, Class: LOC, Start: 86, End: 88, Confidence: 1.00
\end{lstlisting}

\section{System Demonstration}

\subsection{Installation} CyNER is available as a python framework in Github. Users can run the following command for installation: 

\begin{lstlisting}[language=Python]
pip install git+https://github.com/aiforsec/CyNER.git
\end{lstlisting}

\subsection{Entity extraction with a pretrained model}
CyNER allows users to use different combinations of models for extracting entities. In Listing \ref{lst:1} we show a model initialized to use only pretrained XLM-RoBERTa-large model. The model can successfully detect the four cybersecurity-related entities in the text - 'Proofpoint' as Organization, 'FluBot' as malware, 'Android' as System, and the hash as Indicator.

In listing \ref{lst:2}, we combine prediction from the same transformer model with the heuristic-based approach. We use priority 'HTFS,' which means prediction from the heuristic-based detection will be prioritized. The only difference from the previous output is that the hash is now detected as SHA256, a subcategory of the Indicator obtained from regular expression matching. Since the output from the transformer overlaps with the detection from the heuristic and has a lower priority, it is omitted from the merged prediction.

In listing \ref{lst:3}, we add the NER model from Flair on top of transformer and heuristic approaches.\footnote{This is the default parameter setting for prediction} Note that we do not use the Spacy model by default, so it is not used for this prediction. Using the Flair NER model allows us to capture two more generic entities - 'German-language' as MISC (Miscellaneous) and 'UK' as LOC (Location).

\subsection{Fine tuning Models} 
Users can fine-tune transformer language models on annotated NER datasets. The dataset needs to be annotated in the CoNLL 2003 as shown below: 
\begin{lstlisting}[language=Python]
Proofpoint	B-Organization
wrote	O
about	O
the	O
DroidJack	B-Malware
RAT	I-Malware
side-loaded	O
with	O
Pokemon	B-System
GO	I-System
. O
\end{lstlisting}

\begin{lstlisting}[frame=single,label={lst:4},language=Python, caption={Finetuning xlm-roberta-large model}]
cfg = {'checkpoint_dir': '.ckpt',
        'dataset': 'dataset/mitre',
        'transformers_model': 'xlm-roberta-large',
        'lr': 5e-6,
        'epochs': 20,
        'max_seq_length': 128}
model = cyner.TransformersNER(cfg)
model.train()

\end{lstlisting}

Users can configure different parameters, including dataset path, pretrained transformer model, learning rate, epochs, maximum sequence length, as shown in Listing \ref{lst:4}.

%% file: sections/conclusion.tex
\section{Conclusion}
In this paper, we have presented a python library for extracting cybersecurity named entities. We have provided pretrained transformer-based NER models trained on annotated threat reports. In addition, we have added a heuristic-based approach for extracting different indicators and incorporated popular NLP libraries that will allow the user to extract more generic entity types. We have added flexible options to enable the user to combine predictions from multiple models with a defined priority order and support finetune language models on annotated NER corpus. We believe this library will benefit cybersecurity researchers and practitioners to analyze cybersecurity threat reports and gather meaningful insights.